\documentclass[a4paper,twocolumn,11pt,accepted=2023-10-19]{quantumarticle}
\pdfoutput=1
\usepackage[utf8]{inputenc}
\usepackage[english]{babel}
\usepackage[T1]{fontenc}
\usepackage{amsmath}
\usepackage{array}
\usepackage{hyperref}

\usepackage{tikz}
\usepackage{lipsum}
\usepackage[numbers,sort&compress]{natbib}
\usepackage {bookmark}

\begin{document}

\title{Combinatorial optimization solving by coherent Ising machines based on spiking neural networks}

\author{Bo Lu}
 \affiliation{School of Artificial Intelligence, Beijing Normal University, Beijing 100875, China}

\author{Yong-Pan Gao}%
\affiliation{%
School of Electronics Engineering, Beijing University of Posts and Telecommunications, Beijing 100876, China
}%


\author{Kai Wen}
\affiliation{
 Beijing QBoson Quantum Technology Co., Ltd., Beijing 100015, China
}%

\author{Chuan Wang}
 \email{wangchuan@bnu.edu.cn}
\affiliation{School of Artificial Intelligence, Beijing Normal University, Beijing 100875, China}%

\maketitle

\begin{abstract}
Spiking neural network is a  kind of neuromorphic computing that is believed to improve the level of intelligence and provide advantages for quantum computing. In this work, we address this issue by designing an optical spiking neural network and find that it can be used to accelerate the speed of computation, especially on combinatorial optimization problems. Here the spiking neural network is constructed by the antisymmetrically coupled degenerate optical parametric oscillator pulses and dissipative pulses.  A nonlinear transfer function is chosen to mitigate amplitude inhomogeneities and destabilize the resulting local minima according to the dynamical behavior of spiking neurons. It is numerically shown that the spiking neural network-coherent Ising machines have excellent performance on combinatorial optimization problems, which is expected to offer new applications for neural computing and optical computing.
\end{abstract}

\section{Introduction}

Combinatorial optimization problems are of great importance in various fields of computer science. Generally, the combinatorial optimization problem belongs to the non-deterministic polynomial time (NP)-hard complexity class. The effective selection of the best combination among candidates is required in solving the combinatorial optimization problems which aims to maximize the gains or minimize the losses. During the past decades, they have been ubiquitous in various fields, such as drug design \cite{kitchen2004docking}, financial management \cite{soler2017survey}, and circuit design \cite{barahona1988application}.

The computation time of these problems on conventional computers tends to increase exponentially with the size of the problem due to the explosion in the number of combinations \cite{arora2009computational}.  The problem in the complexity class NP could be formulated as an Ising problem with only polynomial overhead, and the loss functions can be mapped to the Ising model to complete the search from the initial state to the ground state, which has been implemented by several physical systems such as quantum annealing \cite{kadowaki1998quantum,das2008colloquium}, optical oscillators \cite{wang2013coherent}, nanomagnetic-net arrays \cite{sutton2017intrinsic,saccone2022direct}.

Among them, oscillator-based coherent Ising machines (CIM) are widely studied due to the advantages \cite{hamerly2019experimental} of high speed, programmability, and parallel search. Each Ising spin in the CIM can be encoded as the phase $\phi_i$ of the light pulse in an optical mode. Employing the degenerate optical parametric oscillation, the phase values are enforced as the binary spin values as $\phi_i=0, \pi$. During computation, the optical gain yields oscillation of the pluses either in-phase or out-of-phase respective to the pump light. Meanwhile, the oscillators are coupled with a tunable coupling constant to form a network. The oscillation network can be implemented by various ways, such as the optical oscillators \cite{marandi2014network,inagaki2016large,mcmahon2016fully,inagaki2016coherent,honjo2021100}, opto-electronic oscillators \cite{Fabian2019A,cen2020microwave}, spatial phase modulators \cite{pierangeli2019large,fang2021experimental,sun2022quadrature}, and integrated photonic chips \cite{okawachi2020demonstration}. 

As a hardware solver, CIM has excellent performance on small-scale combinatorial optimization problems \cite{hamerly2019experimental,mcmahon2016fully,wang2013coherent}, but on large-scale problems, an important technicality would arise in the model which is that the amplitude of the oscillators is not equal. This phenomenon is usually called the amplitude heterogeneity which prevents CIM from being correctly mapped to the Ising model ~\cite{leleu2017combinatorial}, as shown in Fig.1(a). The spin configuration of the ground state of the Ising model differs from that of the CIM. During the calculation of CIM, the energy will be frozen in the excited state of the Ising model, as shown in Fig.1(b). To address the problem of amplitude heterogeneity, there are several methods to fix the problem by adding a feedback mechanism to reduce the inequality of the amplitude,  such as the additional uncorrelated driving signals ~\cite{leleu2017combinatorial}, nonlinear functions \cite{bohm2021order, reifenstein2021coherent} and correcting the amplitude by the error signal \cite{leleu2019destabilization,kako2020coherent, leleu2021scaling}. 
Additionally, the exploration of "dimensional lifting" in Ising machines, particularly those based on multidimensional spins\cite{calvanese2022multidimensional}, has also demonstrated significant enhancements in computational performance.



In addition, we turn our attention to the spiking neural network (SNN) which has been simulated on the CIM based on the DOPOs neural network, and the collective behavior of neural clusters has been studied \cite{inagaki2021collective,lu2023speed}. The spiking neurons are a special type of neuron that performs signal processing in the brain. They are considered as the next generation of neural architectures and have been extensively studied in artificial intelligence recently \cite{maass1997networks,brette2007simulation,izhikevich2003simple}. For example, it can imitate the biological nervous system to produce more intelligent systems, such as brain-like learning \cite{mnih2015human,yu2012pattern}. Furthermore,  there are complex nonlinear dynamics and chaotic phenomena in the spiking neurons \cite{rodriguez1996statistical,harish2015asynchronous}, which hopefully achieve the instability of local minima where the computation of CIMs is frozen.

In this work, we propose a new architecture of CIM, based on a spiking neuron network composed of DOPO pulses and dissipative pulses. We find that it is able to achieve amplitude heterogeneity and also freeze in local minima during computing for combinatorial optimization problems. By controlling the dissipative parameters of spiking neurons composed of antisymmetrically coupled DOPO and dissipative pulses, the rate and direction of contraction of the phase space volumes can be controlled to bring the system out of the frozen state. More generally,  we find the existence of neurons is similar to the dynamic behavior of class-II neurons. An SNN-CIM algorithm is proposed to achieve the instability of the local minimum by adjusting the parameters, and the performance is compared with the algorithm that only uses the nonlinear filter function. The results show that SNN-CIM has excellent performance and has more potential on complex graph structures.

\section{SNN-CIM model}

\begin{figure*}[htbp]
\centering
\includegraphics[width=1\linewidth]{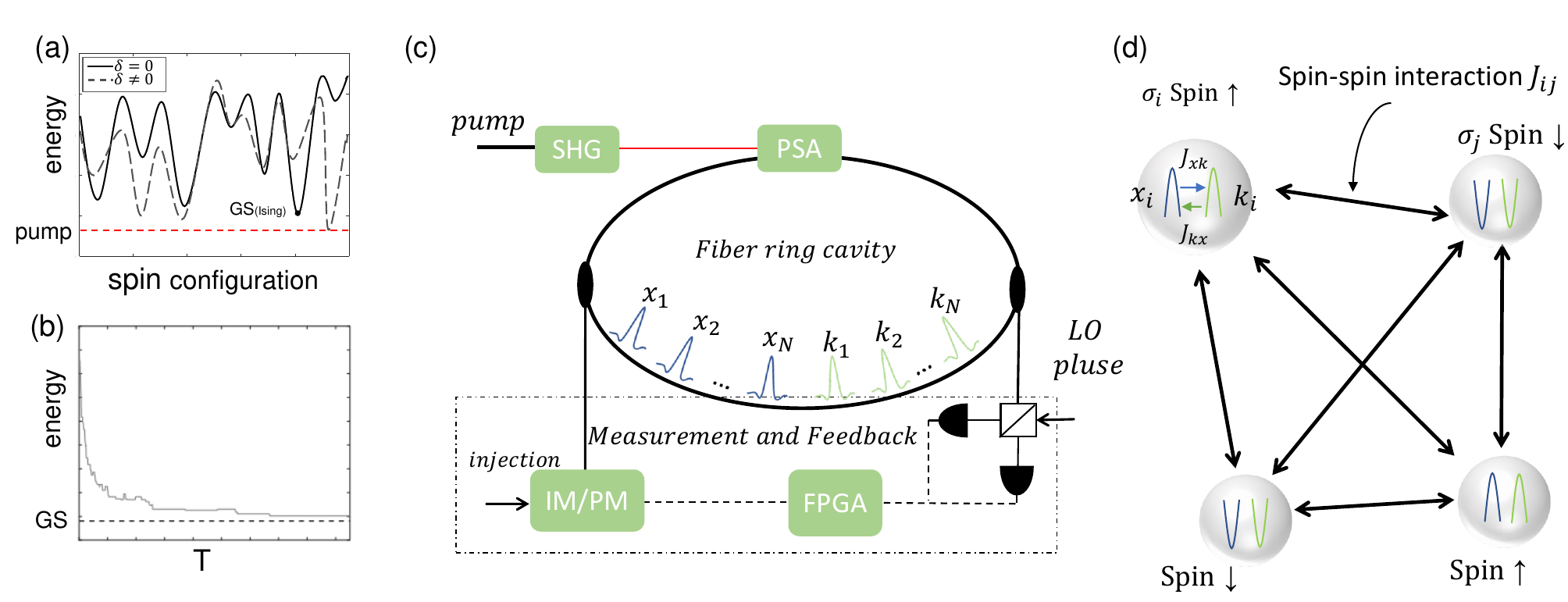}
\caption{Schematic diagram of SNN-CIM. a. Variation of energies with spin configuration for homogeneous and inhomogeneous spin amplitudes. The red line is the pump intensity, which does not excite the correct ground state spin configuration. This figure is for illustration purposes only. b. The time evolution of the energy, due to the amplitude heterogeneity, is frozen in the excited state. The dashed line is the ground state energy. c. Optical implementation of SNN-CIM. The pump is turned on to generate type-x DOPO, and when it is turned off, the modulation generates type-k pulses. d. The type-x DOPO antisymmetrically coupled and dissipated type-k pulses constitute spiking neurons, and the connections between spiking neurons constitute a spiking neural network. IM: intensity modulator, PM: phase modulator, SHG: second harmonic generation, PSA: phase--ensitive amplification, FPGA: field programmable gate array.}\label{fig1}
\end{figure*}

The model of SNN-CIM we use can be summarized in Fig.1(c). The system with correction of amplitude heterogeneity can be implemented using a CIM with a measurement-and-feedback (MFB) structure composed of DOPO pulses. The pulses for computing are represented by the optical SNN pulses	which propagate cyclically in the fiber-ring cavity. For each SNN pulses, it can be divided into two types of pulses by controlling the pump, one is generated by the optical parametric oscillation with nonlinear gain process. It corresponds to the case where the pump is turned on, denoted as $x$, and the other pulse is the injected optical pulse with a dissipative process only, corresponding to the case where the pump is turned off, denoted as $k$. 

A spiking neuron consists of both a type-$x$ pulse and a type-$k$ pulse coupled to each other with coupling strengths of $J_{xk}$ and $J_{kx}$ $(J_{kx}=-J_{xk})$, respectively. The spiking neurons can be connected to each other, as shown in Fig.1(d). Based on this principle, the dynamical behavior of the $i$th neuron can be expressed as:
\begin{equation}
\begin{split}
&\frac{dx_i}{dt}=\alpha x_i-x_i^3+J_{xk}\beta k_i+I_{ext},\\
&\frac{dk_i}{dt}=-\beta k_i+J_{kx}x_i,\\
&I_{ext} = tanh(\epsilon\Sigma_jJ_{ij}x_j).
\end{split}
\end{equation}
 where $x_i$ is in-phase components of DOPO amplitudes, $(k_i)$ is dissipative pulse amplitude. $\alpha$ denotes the nonlinear gain generated by the pump,and  $\beta$ is the inherent dissipation of the system. $J_{ij}$ represents the connection between the $i$th and the $j$th neuron, and $\epsilon$ is the strenght of interaction. 
  
In the case of the power above the threshold, the DOPO pulses undergo spontaneous symmetry breaking. Then, the potential energies of the spiking neurons can be represented by the positive and negative amplitudes of the type-$x$ pulses. Here, the amplitude of each pulse circulating through the fiber loop is measured by the photon detector and recorded by the field programmable gate array (FPGA) module. The measurement results and the coupling matrix are multiplied in the FPGA, filtered by the nonlinear function, and output to the IM and PM to modulate the injected optical pulse. The injected optical pulses are coupled with the optical pulse in the fiber-ring cavity. In Ref.\cite{bohm2021order}, the authors analyzed the nonlinear function filtering has a certain probability of correcting the CIM amplitude heterogeneity.  According to the constraints of the combinatorial optimization problems, the connection of the neurons in the neural network can be regarded as an Ising model that evolves a spin configuration into the ground state.

\section{ The dynamics of the spiking neurons}

\begin{figure*}[htbp]
\centering
\includegraphics[width=1\linewidth]{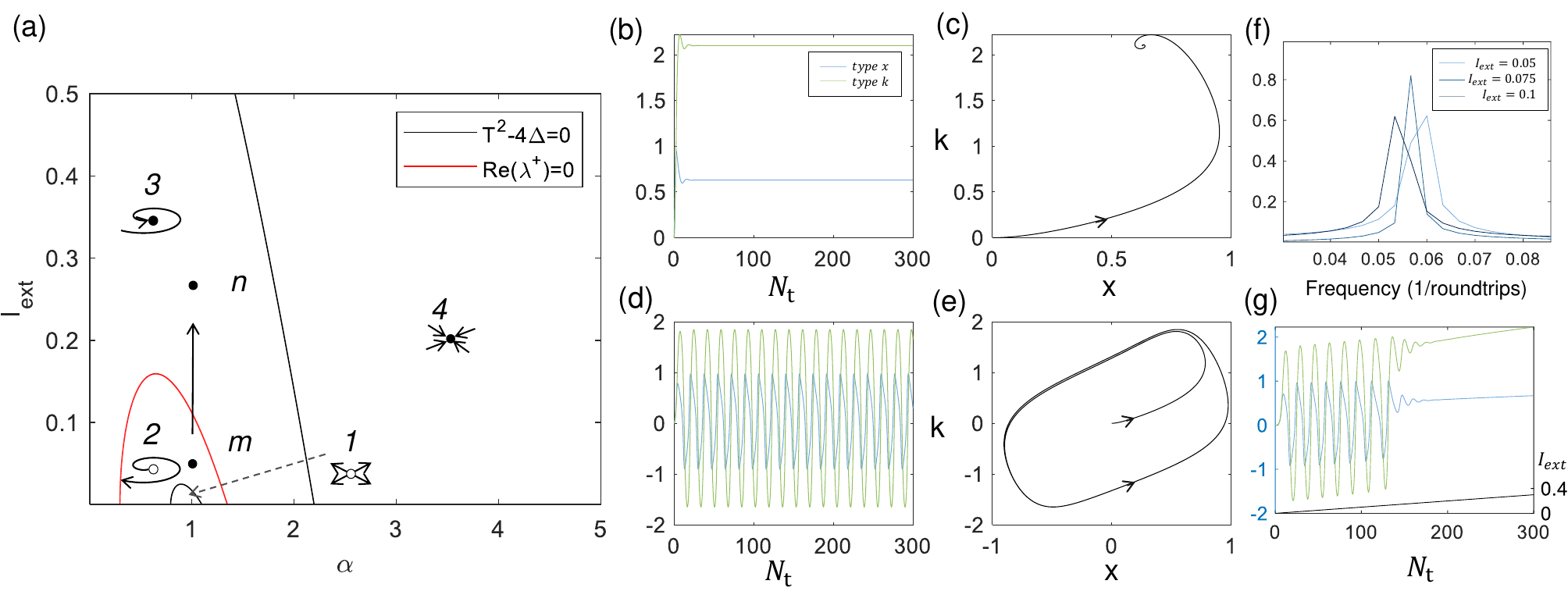}
\caption{Dynamic behavior of spiking neurons. a. The effect of pump strength $\alpha$ and $I_{ext}$ on the stability of the fixed point. The eigenvalues in region 1 and region 4 are real numbers, one has a positive number and the other is all negative. The eigenvalues of region 2 and region 3 are complex conjugates, one is the positive real part and the other is the negative real part. $\beta =0.3$. b. and c. The time evolution and phase space trajectories of the type-$x$ DOPO pulse and type-$k$ pulse at point n in (a), $\alpha=1$ and $I_{ext}=0.25$. d. and e. Corresponding to m points, $\alpha=1$ and $I_{ext}=0.05$. f. Fourier transform of the time evolution of amplitude at different $I_{ext}$. g. The time evolution of the type-$x$ DOPO pulse and type-$k$ pulse with the increase of $I_{ext}$.}\label{fig2}
\end{figure*}

According to Eq.(1), we find the dynamic properties of a spiking neuron depend on the linear stability of its fixed point. The fixed points under the steady-state solution of Eq.(1) should satisfy the relation as:
\begin{equation}
\begin{split}
&0=-\beta k_i+x_i,\\
&0=\alpha x_i-x_i^3+\beta k_i+I_{ext}.
\end{split}
\end{equation}
The solution satisfying Eq.(2) is denoted as $x_0$ and the linear stability of a fixed point $(x_0,x_0/\beta)$ can be obtained by the eigenvalues of its Jacobian matrix:
\begin{equation}
J_i=\left[ \begin{array}{cc}
\alpha-3x_0^2&-\beta\\
1&-\beta\\
\end{array} \right ].
\end{equation}
Here the eigenvalues $\lambda_i^{\pm}$ of the Jacobian matrix are given as follows:
\begin{equation}
\lambda_i^{\pm}=\frac{1}{2}(T\pm\sqrt{T^2-4\Delta}),
\end{equation}
where $\Delta$ and $T$ are the determinant and trace of the Jacobian matrix, respectively. Then, we have $\Delta=J_{xx}J_{kk}-J_{xk}J_{kx}=(3x_0^2-\alpha+1)\beta$ and $T=J_{xx}+J_{kk}=\alpha-3x_0^2-\beta$. 
Numerical simulations give the fixed-point linear $((x_0,x_0/\beta))$ stability as a function of $I_{ext}$ and the pump strength $\alpha$ when $\beta = 0.3$, as shown in Fig.2(a). The stability of those fixed points depends on $Re(\lambda_i^{+})$, these fixed points become unstable when $Re(\lambda_i^{+})>0$. However, the eigenvalues would become complex conjugated when $T^2-4\Delta<0$.
Hence, we mark $Re(\lambda_i^{+})=0$ and $T^2-4\Delta=0$ in Fig.2(a), respectively. Then, the time evolution and the oscillation phase space of those fixed points in regions 3 and 2 are shown in Fig.2(b)(c) and Fig.2(d)(e), respectively. Under a small disturbance of $I_{ext}$, the system reciprocates on a limit cycle. And the system will shrink to a stable focus under a large $I_{ext}$. 

The spiking behavior of single neurons can be achieved by controlling the parameter $I_{ext}$. When $I_{ext}$ increases, the amplitude of neuron pulses gradually decreases. However, for the increment of $I_{ext}$ to a certain value, the neuron pulses will no longer appear, as shown in Fig.2(g). 
This process is consistent with the evolution direction of the arrow indicated by Fig.2(a), that is, the system appears Andronov-Hopf (AH) bifurcation with the increase of $I_{ext}$, which makes the fixed points change from an unstable limit cycle to a stable focused point. This shows that the optical pulse network has the properties of the class-II neurons by anti-symmetrically coupling a pair of nonlinear gain DOPO pulse and dissipation optical pulse.

In addition, as shown in Fig.2(f), we find a smaller $I_{ext}$ could exhibit a larger firing rate by taking a Fourier transform of the temporal evolution of neurons under different $I_{ext}$. For the Ising model with antiferromagnetic coupling, the evolution of the energy $E_{Ising}=-\Sigma_{ij}\sigma_i\sigma_j/2$ is accompanied by an amplification of the coupling term. 
In this model, $I_{ext}$ corresponds to the coupling term on the $i$th spin, that is, the firing rate decreases ($I_{ext}$ increases) during the evolution of the Ising model. 
Moreover, the cluster neurons can also be synchronized to a low firing rate. 

\begin{figure}[htbp]
	\centering
	\includegraphics[width=1\linewidth]{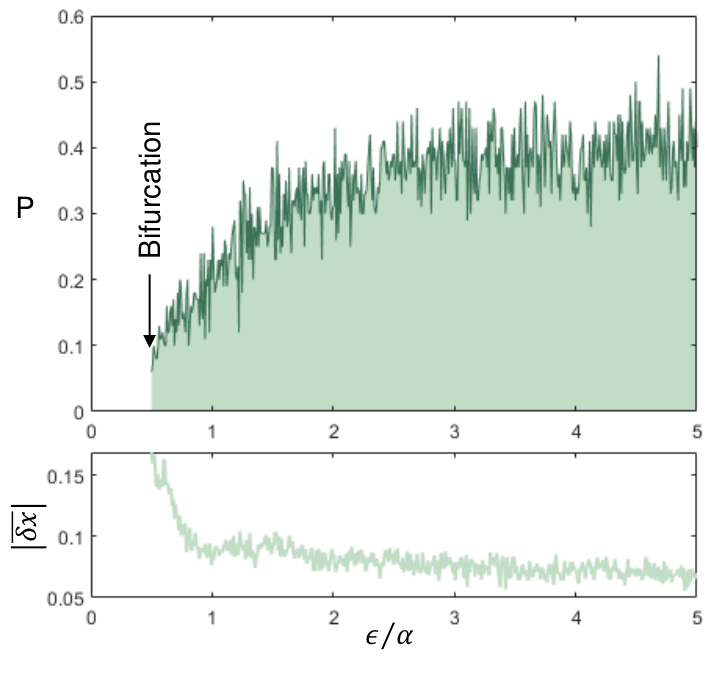}
	\caption{Relation between amplitude heterogeneity correction and coupling parameters.
	The top figure shows the variation of the success rate $P$ of solving the Max-Cut problem of the N=25 graph with the coupling strength $\epsilon/\alpha$ (normalized by the pump strength $\alpha$).
	The bottom figure is the correction of amplitude inhomogeneity $|\bar{\delta x}|$ under different coupling strength $\epsilon/\alpha$. Other parameters $\alpha = 1.0$.
	}\label{fig3}
\end{figure}


To foster a more profound comprehension of the amplitude heterogeneity correction facilitated by the CIM based on spiking neural networks, our study focuses on a simplified graph with dimensions $N=25$ nodes and $E=75$ edges, selected as the testbed. We meticulously examined the success rate of solutions denoted as $P$ and the mean absolute error of the amplitude of different nodes represented as $|\bar{\delta x}|$ across varying coupling strengths $\epsilon/\alpha$, as shown in Fig.3.
We can find that as the coupling strength increases, the system energy starts to decrease with the increase of $I_{ext}$, so the success rate gradually increases. When the $tah$ function response is saturated, the probability gradually stabilizes.
Moreover, as the coupling strength increases, the difference in amplitude increases in the conventional CIM\cite{bohm2021order}, but in our scheme, the nonlinear function causes the difference to disappear gradually.
Furthermore, there is a correlation between the ability to find the ground state and the amount of amplitude inhomogeneity. As the inhomogeneity of the magnitude decreases, the mapped Ising model gets closer to the target Hamiltonian, increasing the ability to find the optimal solution.

\section{Instability of the local Minimum and SNN-CIM Algorithms}

\begin{figure*}[htbp]
\centering
\includegraphics[width=1\linewidth]{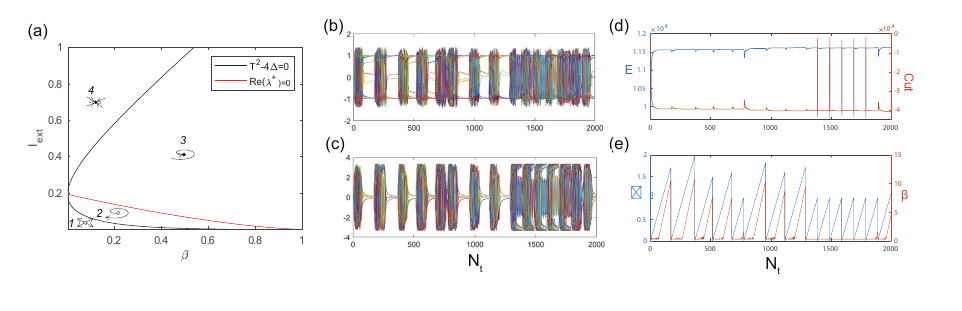}
\caption{SNN-CIM solves the MAX-CUT problem of G1. a. The effect of $\beta$ and $I_{ext}$ on the stability of the fixed point. The eigenvalues in region 1 and region 4 are real numbers, one has a positive number and the other is all negative. The eigenvalues of region 2 and region 3 are complex conjugates, one is the positive real part and the other is the negative real part. $\alpha =1.0$. b. Evolution of type-$x$ DOPO pulse amplitude with calculation time. c. Evolution of type-$k$ pulse amplitude with calculation time. d. Energy and number of cuts as a function of computation time. e. The coupling strengths $\epsilon$ and the dissipation $\beta$ as a function of computation time.}\label{fig3}
\end{figure*}

Due to the heterogeneity of amplitudes in the computation of CIM, there are differences of the ground state energy configuration between $\delta=0$ and $\delta\neq0$, as shown in Fig. 1(a). Meanwhile, the Ising machines will be frozen at the local minimum and cannot evolve towards the global optimum, as shown in Fig. 1(b),  which is observed by Ref. \cite{bohm2018understanding} on the CIM implemented by the DOPOs pulses network.

Similar to the analysis of the linear stability of the pump and $I_{ext}$ on the fixed points, Fig.4(a) shows the effect of dissipation $\beta$ and $I_{ext}$ on the linear stability of those fixed points. 
Those fixed points that freeze at local minima tend to be located in region 3.  The stability of the local minima can be influenced to escape from the frozen local minima by adjusting the dissipation and coupling strength.
To overcome the local minima and make the system to escape from the frozen state,  we choose $Re(\lambda_i^+)>0$ by reducing the dissipation $\beta$ and coupling strength $\epsilon$ which forces these fixed points into region 2 to destabilized. Then, the system continues to search for the ground state with lower energy by gradually increasing the coupling strength. 

We design an SNN-CIM algorithm according to Eq.(1), which evolves under the pumping intensity $\alpha=1$. 
We reduce the energy of the system by increasing the coupling strength $\epsilon$ and dissipation coefficient $\beta$, so that the system transitions from oscillation to stable bifurcation.
After each iteration, we record the energy of the system at this moment $E_t$ and compare it with the lowest energy before the iteration.
 If the energy is higher than the minimum energy at this time, it means that the system evolution has entered a locally optimal solution. We need to reset the lower coupling strength and dissipation coefficient to make the system enter the unstable region (1 or 2) to re-oscillate to search for the optimal solution.
After repeating the above operations, after the total calculation time $N_t$, SNN-CIM can give an optimal solution or an acceptable suboptimal solution. 
The selection of parameters $\beta$ and $\epsilon$ can be adjusted according to the fixed point stability of Fig.4(a).
Here, we establish a step size of $h=0.05$, with an initial value of $\epsilon_0=0$ for the coupling strength $\epsilon$ and $\beta_0=0.3$ for the dissipation strength $\beta$. During each iteration, we increment $\epsilon$ by $0.01h$ and $\beta$ by $0.1h$.  In the event that the energy $E_t$ does not exhibit a reduction during the iteration process, the parameters  $\epsilon$ and $\beta$ are reset to their respective initial values.

To give an example of the algorithm, we apply the SNN-CIM algorithm to the MAX-CUT problem of G1. We start with a graph, in which some of the vertices are connected via edges. The aim of the MAX-CUT problem is to group the vertices into two types with the number of edges as large as possible, whose mathematical expression is generally considered to be equivalent to the Ising model as
\begin{equation}
edges = \frac{1}{4}(\sum_{ij}J_{ij}-\sum_{ij}J_{ij}\sigma_i\sigma_j).
\end{equation}

In Fig.4(b), the time evolution of the type-$x$ DOPO and type-$k$ pulses, the energy, and the number of cuts of the system are given. Specifically, if the energy is frozen in a local minima state, we tend to reduce $\beta$ which makes the system unstable. Then, there is fluctuation of the energy and the system continues to search for the lower energy state. Finally, if the system energy is higher than the lowest energy before this moment, as in Fig.4, the optimal solutions could be obtained as $edges =11624$ near $N_t=1350$, after that the system will be unstable until another lower energy is searched.

\section{The performance of SNN-CIM Algorithm on MAX-CUT Problem}

Here in this part, we describe the performance of the SNN-CIM algorithm on the Max-Cut problem and then compare it with other types of computing devices such as those based on simulated annealing and conventional CIM. We discuss the performance of these investigated devices on different sizes of the Ising problems. 

We chose the simulated annealing (SA) algorithm and the SNN-CIM algorithm on the G1-G21 graph of 800 nodes, and the G22-G42 graph of 2000 nodes, and also compared the performance with the conventional CIM and the CIM with sigmoid function optimization in Ref.\cite{bohm2021order}. 

\begin{figure*}[htbp]
\centering
\includegraphics[width=1\linewidth]{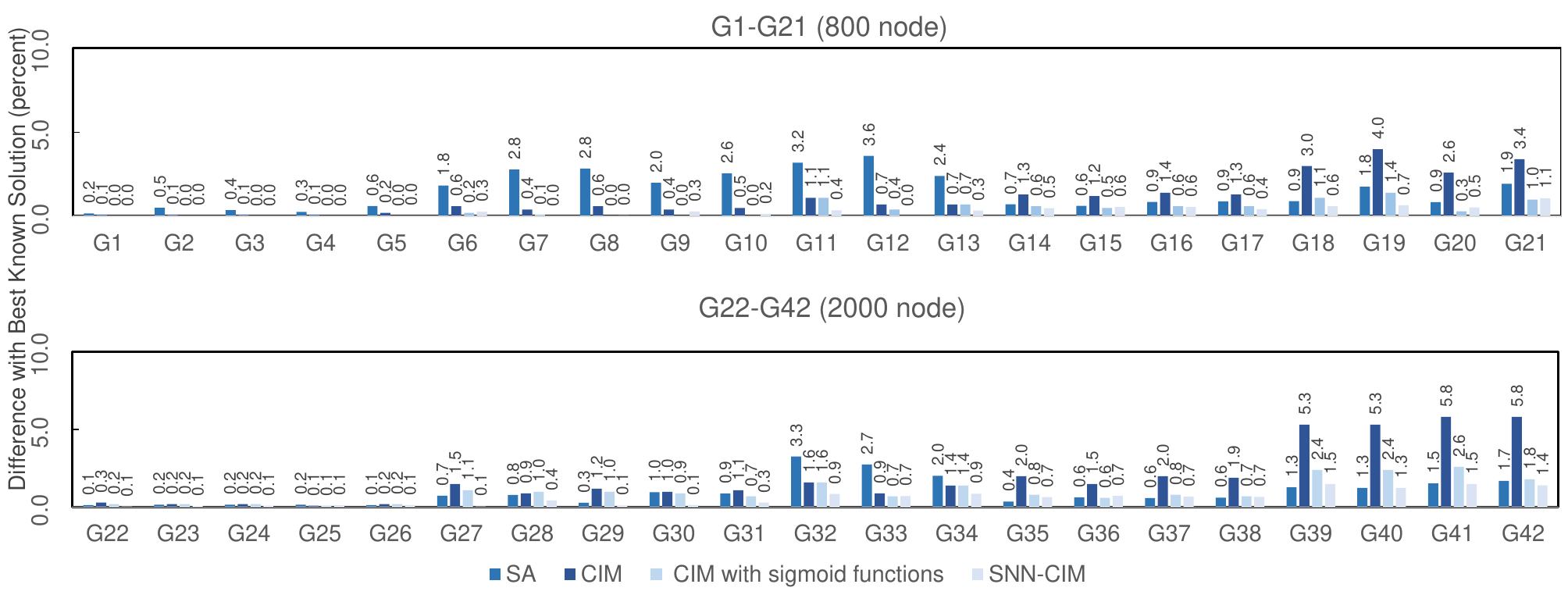}
\caption{Computational performance of the SNN-CIM algorithm on the MAX-CUT problem for a given problem. a. Relative error of the optimal value of simulated annealing algorithm, CIM, CIM with sigmoid function, and SNN-CIM for 50 runs on the MAX-CUT problem of a graph with 800 nodes. b. More complex 2000 node situation.}\label{fig4}
\end{figure*}

The performance of the computational results are shown in Fig.5. Here we choose the relative error $\Delta C=100(1-C/C_{opt})$ of the optimal value during 50 runs as the result, and $C_{opt})$ denotes the best-known value reported in the literature.  In the 800 nodes problem, CIM with sigmoid function and SNN-CIM often find the solutions very close to the optimal value, while the performance of the SA and CIM are relatively poor. And, the mean values of the relative errors of SNN-CIM and CIM with sigmoid function in the optimal value of G1-G21 are $0.3\%$ and $0.4\%$, respectively. The performance of SNN-CIM is slightly better than that of CIM with a sigmoid function. Under the more complex 2000-node problem, SNN-CIM is significantly closer to the optimal value than the other three algorithms.
In addition, for more complex problems with random connectivity and non-uniform node degrees (G39,G40,G41,G42), SNN-CIM with unstable local optima can also achieve better computational performance. This shows that SNN-CIM will have more potential in more complex combinatorial optimization problems.

\renewcommand\arraystretch{1.5} 
\begin{table*}[htb] 
		\setlength{\abovecaptionskip}{0cm} 
		\setlength{\belowcaptionskip}{0.2cm}
	\centering
	\caption{\label{tab:l1} Computational performance of SA, CIM, and SNN-CIM on GSET. TTS is a time-to-solution of reaching 98 percent of the best solution. The optimal solutions for SA, CIM, and SNN-CIM are obtained under 10, 50, and 50 runs, respectively. Values in parentheses indicate that the results of $TTS_{98}$ cannot be obtained, and are replaced by $TTS_{95}$. The best maximum cuts are known from Ref.\cite{ma2017multiple}} 
	\begin{tabular}{p{0.3cm}<{\centering}p{0.5cm}<{\centering}p{1cm}<{\centering}p{1cm}<{\centering}p{1cm}<{\centering}p{1cm}<{\centering}p{1cm}<{\centering}p{1cm}<{\centering}p{2cm}<{\centering}p{2cm}<{\centering}p{2cm}<{\centering}} 
		\hline
		ID& N & $J_{ij}$ & Edge & $C_{opt}$ & $C_{SA}$ & $C_{CIM}$ & $C_{SC}$ & $TTS_{SA}(s)$ & $TTS_{CIM}(s)$ & $TTS_{SC}(s)$\\ 
		 \hline
		1&800&{$\{0,1\}$}&19176&11624&11604&11597&11624&16.1514&3.8065&0.0373 \\ 
		6&800&{$\{0,\pm1\}$}&19176&2178&2138&2149&2172&20.4624&4.0332&1.3423 \\ 
		11&800&{$\{0,\pm1\}$}&1600&564&546&550&562&(8.46337)&(0.4107)&0.4775 \\ 
		14&800&{$\{0,1\}$}&4694&3064&3042&2993&3049&11.9894&(1.8703)&0.0786 \\ 
		18&800&{$\{0,\pm1\}$}&4694&992&983&942&986&20.4379&(8.418)&1.52 \\ 
		22&2000&{$\{0,1\}$}&19990&13359&13340&13279&13352&2169.69&14.048& 0.9654\\ 
		27&2000&{$\{0,\pm1\}$}&19990&3341&3316&3271&3339&2760.23&(23.568)& 27.768\\ 
		32&2000&{$\{0,\pm1\}$}&4000&1410&1364&1364&1398&(2909.38)&(7.1411)&32.101\\ 
		35&2000&{$\{0,1\}$}&11778&7684&7656&7448&7634&1644.46&(14.806)&1.2013 \\ 
		\hline
	\end{tabular}
\end{table*}
The computational performance of the SNN-CIM algorithm on GSET is also shown in Table 1, which is compared with SA and CIM. We note that the algorithm of SNN-CIM has an unparalleled advantage in computing time to find suboptimal but adequate solutions for the application, both in the time and accuracy of the solutions. Based on the above analysis, our proposed SNN-CIM has excellent performance on combinatorial optimization problems.

\section{Conclusion}

In summary, we propose a spiking neural network consisting of DOPO pulses and dissipative pulses. We analyze the fixed points stability of a single spiking neuron and investigate the dynamical behavior under different $I_{ext}$ through numerical simulations.  The neuron is in an unstable firing state when the smaller coupling term acts as $I_{ext}$, and the neuron pulse disappears with the increase of the coupling term, corresponding to the AH bifurcation process, simulating the class-II neurons. Based on the properties of spiking neurons, an SNN-CIM algorithm for solving combinatorial optimization problems is proposed. 
By increasing the coupling strength, the neuron cluster in the firing state is bifurcated to search for the ground state of the Ising model, and a nonlinear filter function is used to eliminate the amplitude heterogeneity. For the local minimum point frozen in the excited state during the filtering process, it can be destabilized by adjusting the dissipation parameter $\beta$, so as to continuously search for the ground state. We also compare the performance of the SNN-CIM algorithm on combinatorial optimization problems with SA, CIM, and CIM with sigmoid function. The results show that SNN-CIM has excellent performance on combinatorial optimization problems, and has more potential on more complex problems with random connectivity and non-uniform node degrees.

In addition, the model can be implemented by adding dissipative pulses to the existing CIM based on the fiber ring cavity, which is expected to improve its computational performance and solve the problem of freezing in the excited state during the calculation process. By using integrated photonics direct coupling and high repetition frequency lasers, it is expected to increase the pulse firing rate and the number of neurons, improve computing power, and provide a better simulation platform for spiking neural networks.
Finally, the use of the sigmoid function for the coupled term response in this model, which is widely used in neural networks, creates an interesting connection between CIM and neuromorphic computing, with the potential to excite the research in the perceptron and pattern recognition of fiber-rings cavity based CIM.

\section{Acknowledgement}

This research was funded by the National Natural Science Foundation of China (Grants Nos. 62131002, 62371050), and the Fundamental Research Funds for the Central Universities (BNU).

\section{Data availability}

The GSET instances used for the tests are available at https://web.stanford.edu/~yyye/yyye/Gset/. The SA algorithm temperature adjustment is based on a logarithmic function, $T=T_0log_2(1+t_0/t)$. All programs are written by Matlab and run on intel i7-10700F 2.9GHz.

\bibliographystyle{quantum}
\bibliography{quantumview-template}

\end{document}